\documentclass[pra,twocolumn,showpacs,a4paper]{revtex4}
\usepackage{amssymb}
\usepackage{epsfig}
\usepackage{amsmath}
\usepackage{graphicx}
\usepackage{bm}
\usepackage{times}
\usepackage{txfonts}
\usepackage{color}
\usepackage{hyperref}
\usepackage{comment}

\begin{document}

\title{Phase-noise induced limitations on cooling and coherent evolution in
opto-mechanical systems}
\author{P. Rabl}
\affiliation{ITAMP, Harvard-Smithsonian Center for Astrophysics, Cambridge, Massachusetts
02138, USA}
\author{C. Genes, K. Hammerer}
\affiliation{Institute for Theoretical Physics, University of Innsbruck, and \\
Institute for Quantum Optics and Quantum Information of the Austrian Academy
of Sciences, 6020 Innsbruck, Austria.}
\author{M. Aspelmeyer\footnote{Permanent address: Faculty of Physics, Boltzmanngasse 5, University of Vienna, 1090 Vienna, Austria}}
\affiliation{Institute for Quantum Optics and Quantum Information, Austrian Academy of
Sciences, Boltzmanngasse 3, 1090 Wien, Austria}
\date{\today}

\begin{abstract}
We present a detailed theoretical discussion of the effects of ubiquitous laser noise on cooling and the coherent dynamics in opto-mechanical systems. Phase fluctuations of the driving laser induce modulations of the linearized opto-mechanical coupling as well as a fluctuating force on the mirror due to variations of the mean cavity intensity. We first evaluate the influence of both effects on cavity cooling and find that for a small laser linewidth the dominant heating mechanism arises from intensity fluctuations.  The resulting limit on the final occupation number scales linearly with the cavity intensity both under weak and strong coupling conditions. For the
strong coupling regime, we also determine the effect of phase noise on the
coherent transfer of single excitations between the cavity and the
mechanical resonator and obtain a similar conclusion. Our results show that 
conditions for optical ground state cooling and coherent operations are experimentally feasible and thus laser phase noise does pose a challenge but not a stringent limitation for
opto-mechanical systems.
\end{abstract}

\pacs{42.50.Lc,          42.50.Wk,           07.10.Cm           }
\maketitle

Over the past years tremendous experimental progress with opto-mechanical
devices~\cite{Kippenberg2008,Marquardt2009} and analogous systems in the
microwave regime~\cite{TeufelNJP2008,Rocheleau2009} has been made. Many
groups have by now achieved significant cooling of mechanical motion~\cite%
{GiganNAT2006,ArcizetNAT2006,KlecknerNAT2006,CorbittPRL2007,Schliesser2008,ThompsonNAT2008,Wilson2009}
and in several of these setups resolved-sideband conditions have been
demonstrated, which is a necessary
prerequisite for quantum ground state cooling~\cite%
{MarquardtPRL2007,WislonRaePRL2007,GenesPRA2008,YongLiPRB2008}.
Beyond laser cooling, which works well in the regime of weak
opto-mechanical coupling, the recently demonstrated strong coupling regime
\cite{Groblacher2009} might allow for the observation of coherent dynamics
between the cavity field and the vibrating mirror. For example, the transfer
of single photonic excitations to the phonon mode has been suggested \cite%
{Marshall2003,RomeroIsart2009} to prepare and study quantum superpositions
of macroscopic objects.

Despite a steady experimental progress cooling to the ground state and
further the combination of strong coupling conditions with low occupation
numbers have not been achieved so far. The main limitation in current
systems arises from the re-thermalization rate of the mirror $%
\Gamma _{m}=k_{B}T/\hbar Q_{m}$, (where $Q_{m}$ is the mechanical quality
factor and $T$ the temperature of the support), which competes
with optical cooling. However, with lower base temperatures and increasing
mechanical quality factors, mechanical heating can be strongly reduced and
the impact of other imperfections on opto-mechanical systems must be
considered. In particular, it has been argued recently \cite{DiosiPRA2008}
that ubiquitous laser phase fluctuations impose severe limitations on
opto-mechanical cooling schemes and can impede ground state cooling under
current experimental parameters. While effects of laser noise have indeed
been observed~\cite{Schliesser2008}, the laser linewidth requirements $%
\Gamma _{L}\approx 10^{-3}$ Hz for ground state cooling estimated in \cite{DiosiPRA2008} are clearly inconsistent with current experimental
achievements where residual mean occupancies as low as $n_{0}\simeq 30$ \cite{Groeblacher2008,ParkNatPhys2009,SchliesserNatPhys2009} have been achieved
with $\Gamma _{L}\sim 1$ kHz. This discrepancy can be understood from a
suppression of noise at the mechanical frequency $\omega _{m}$~\cite%
{Schliesser2008}, but a rigorous derivation has not been given so far.
Furthermore, potential impairing effects of a finite laser linewidth on
strongly coupled opto-mechanical systems have not been addressed yet.

Given the great interest in opto-mechanical experiments the impact of phase
noise in such systems deserves a thorough theoretical investigation, which
we provide in this work. We here generalize the standard descriptions of
opto-mechanical systems to include laser phase noise with arbitrary frequency noise spectrum $S_{\dot{\phi}}(\Omega )$
and evaluate its influence on
cooling and coherent oscillations. 
The main results of this work are as follows. In accordance with previous predictions~\cite%
{Schliesser2008,DiosiPRA2008} we obtain a lower limit for the final mirror occupation number which is
proportional to the intensity and the noise spectrum $S_{\dot{\phi}}(\omega
_{m})$ at the mechanical resonance frequency $\omega _{m}$.  Surprisingly, this result applies for both the weak
and strong coupling regime and we find that for a given $\Gamma _{m}$ and $g_{0}$
(opto-mechanical coupling \textit{per single photon)}, the condition 
\begin{equation}\label{eq:Cond1}
S_{%
\dot{\phi}}(\omega _{m})<g_{0}^{2}/\Gamma _{m},
\end{equation} 
must be satisfied in order
to achieve ground state cooling. In the strong coupling regime we derive a
similar condition for the observation of coherent oscillations, 
\begin{equation}\label{eq:Cond2}
S_{\dot{\phi%
}}(\omega _{m})<g_{0}^{2}/\kappa,
\end{equation}
where $\kappa $ is the cavity field decay rate. Importantly, we also show
that under relevant conditions decoherence from low frequency phase noise in opto-mechanical systems is negligible. Thus a small laser
linewidth $\Gamma _{L}/\kappa \ll 1$ together with suppression of phase noise at high frequencies is sufficient to enable
ground state cooling and the observation of coherent oscillations. The conditions~\eqref{eq:Cond1}  and~\eqref{eq:Cond2} are
experimentally challenging, but well within reach with state of the art
laser stabilization.

%

In the next section we develop the model for opto-mechanical coupling
including laser noise, closely following the lines of \cite%
{WislonRaePRL2007,WilsonRaeNJP2008}. The results on weak and strong coupling
are derived in Sec.~\ref{sec:weak} and Sec.~\ref{sec:strong}, respectively.
Details of calculations are moved to three appendices.

\section{Model}

We consider a typical opto-mechanical setup of an optical cavity
mode coupled to a micro- or nanomechanical oscillator. We will refer
in the following to the prototype example of an opto-mechanical
system, a Fabry-Perot cavity of length $L$ with one fixed, heavy
mirror (which serves as the input coupler)\ and a light vibrating
micro-mirror. However, our results apply as well to microtoroidal
cavities or to dielectric membranes in a Fabry-Perot cavity. We
restrict our analysis to a single vibrational mode of the mirror of
mass $m$ and frequency $\omega _{m}$ and a single cavity mode of
frequency $\omega _{c}$ that is driven close to resonance by a laser
of frequency $\omega _{l}$. According to \cite{LawPRA1995} the field
couples to the motion of the mirror via the radiation pressure
interaction
and a total Hamiltonian can be written (in a frame rotating at $\omega _{l}$)%
\begin{equation}\label{eq:ham}
H=-\Delta _{0}a^{\dag }a+\omega _{m}b^{\dag }b+g_{0}a^{\dag
}a(b+b^{\dag })-i\left( \mathcal{E}^{\ast
}(t)a-\mathcal{E}(t)a^{\dag }\right) \,.
\end{equation}%
Here $a$, $a^{\dag }$ are the bosonic operators for the cavity mode
such that $[ a,a^{\dagger} ] =1$ and $b$, $b^{\dag }$ for the
mechanical mode $[ b,b^{\dag }] =1$, while $\Delta _{0}=\omega
_{l}-\omega _{c}$. The optomechanical coupling is $g_{0}=\left(
\omega _{c}/L\right) \sqrt{\hbar /\left( 2 m\omega _{m}\right) }$
and $\mathcal{E}(t)$ is the noisy laser field with an absolute value
$|\mathcal{E}(t)|\sim\mathcal{E}_{0} $. While our analysis can be
generalized to arbitrary noise processes we focus in the following
solely on phase noise,
\begin{equation}
\mathcal{E}(t)=\mathcal{E}_{0}e^{i\phi (t)}\,.
\end{equation}
Here $\phi (t)$ is a Gaussian noise process with zero mean which is
characterized by the correlation function
\begin{equation}  \label{eq:NoiseCorrelations}
\left\{ \dot{\phi}(s)\dot{\phi}(s^{\prime })\right\} _{cl}= \int
\frac{d \Omega}{2\pi} S_{\dot \phi} (\Omega)
e^{-i\Omega(s-s^{\prime})}\,,
\end{equation}
where $\left\{ O\right\} _{cl}$ denotes the average over different
noise realizations. The frequency noise spectrum $S_{\dot
\phi}(\Omega)$ is specific for each experimental setup. For
concreteness we will consider below a simplified noise model with a
noise spectrum and correlation function,
\begin{align}  \label{eq:Noise}
S_{\dot \phi}(\Omega)&= \frac{2 \Gamma_L \gamma_c^2}{\gamma_c^2+\Omega^2}%
, & \left\{ \dot{\phi}(s)\dot{\phi}(s^{\prime })\right\}
_{cl}&=\Gamma_L \gamma_c e^{-\gamma_c|s-s^{\prime}|},
\end{align}
respectively. Here $\Gamma_L $ can be identified with the laser
linewidth which is typically known. The additional parameter $\gamma
_{c}^{-1}$ characterizes a finite correlation time of the underlying
noise process and leads to a suppression of noise at high
frequencies. In the limit $\gamma _{c}\rightarrow \infty $ we obtain
the $\delta $-correlated noise model used in \cite{DiosiPRA2008}.

In addition to the classical driving field, the cavity field is
coupled to the electromagnetic modes of the environment. We describe
the resulting dissipative dynamics by a master equation
\begin{equation}  \label{Eq:MEQ}
\dot{\rho}=-i[H,\rho ]+\mathcal{L}_f(\rho),
\end{equation}%
where $\mathcal{L}_f(\rho)=\kappa (2a\rho a^{\dag }-a^{\dag
}a\rho-\rho a^{\dag }a)$ and $\kappa$ is the cavity field decay
rate. We made use of the fact that for optical experiments even at
room temperature the thermal
photon number is effectively zero~\cite{Thermalnoise}. In principle Eq. %
\eqref{Eq:MEQ} should also contain the term describing coupling of
the mechanical system to its thermal environment. The influence of thermal noise on opto-mechanical cooling schemes has been studied 
in previous works~\cite%
{MarquardtPRL2007,WislonRaePRL2007,GenesPRA2008,YongLiPRB2008}. To focus on the effects of laser noise only we here consider the regime of efficient laser cooling where
thermal noise is already suppressed and the dynamics is well described by the master equation~\eqref{Eq:MEQ}.

\subsection*{Linearized opto-mechanical coupling}

The radiation pressure from the cavity field leads to a mean
displacement of the mirror and we are only interested in
fluctuations around this shifted equilibrium position. We therefore
perform a unitary transformation $b\rightarrow \beta +b$ where
$\beta=-g_{0}n_{ph}/\omega _{m} $ is the mean displacement amplitude
and $n_{ph}=\{ \langle a^{\dag }a\rangle _{q}\} _{cl}$ is the mean
cavity photon number averaged over both quantum and classical
fluctuations. In addition, we perform another unitary displacement
operation for the intracavity field $a\rightarrow \alpha (t)+a$
where%
\begin{equation}  \label{Eq:alpha}
\alpha (t)=\int_{-\infty }^{t}ds\,e^{-(-i\Delta +\kappa )(t-s)}\mathcal{E}%
(s)\,,
\end{equation}%
is the classical part of the stochastically driven cavity field and
$\Delta =\Delta _{0}-2g_{0}\beta $ is the effective detuning. Note
that by choosing this time-dependent displacement amplitude
$\alpha(t)$ instead of a constant one~\cite{DiosiPRA2008} we can unambiguously interpret $a$ as the
quantum field of a non-driven cavity.

From these transformations we obtain the master equation
\begin{equation}  \label{eq:TransMasterEq}
\dot{\rho}=-i[H_{\mathrm{opt}}(t)+g_0\mathcal{N}(t)(b+b^\dag),\rho ]+%
\mathcal{L}_f(\rho),
\end{equation}%
where $\mathcal{N}(t)=|\alpha (t)|^{2}-\{|\alpha(t)|^2\}_{cl}$
measures intensity fluctuations of the mean intracavity field and the
opto-mechanical is now given by%
\begin{equation}  \label{Eq:DispHam}
\begin{split}
H_{\mathrm{opt}}(t)=& \omega _{m}b^{\dag }b-\Delta a^{\dag
}a+g_{0}\left( \alpha ^{\ast }(t)a+\alpha (t)a^{\dag }\right)
(b+b^{\dag })\,.
\end{split}%
\end{equation}%
Note that in Eq.~\eqref{Eq:DispHam} we have already neglected the
nonlinear
term $\sim a^{\dag }a$ which is equivalent to an expansion in $1/\sqrt{n_{ph}%
}$ and well justified by current experimental setups.

From the equation of motion \eqref{eq:TransMasterEq} we conclude
that laser noise contributes two effects: First, it causes a modulation of the opto-mechanical coupling strength $\sim
\alpha (t)$. Second, it induces a stochastic force on the mirror
$\sim \mathcal{N}(t)$, which arises from a conversion of phase to amplitude fluctuations inside the optical resonator.
This heating mechanism has no direct analog in laser cooling in free space 
but a similar effect has been observed with trapped atoms inside a cavity~\cite%
{SavardPRA1997,YePRL1999}. In the following we evaluate the influence
of both contributions on cavity cooling and the coherent evolution
in this system.


\section{Weak coupling regime}\label{sec:weak}

We first consider the weak coupling limit $g_{0}|\alpha(t)|\ll
\kappa $. In this regime the cavity mode acts as a dissipative
channel for the resonator mode which for example can be exploited
for cooling. We are interested in the effects of phase noise on the
mean mirror occupation number in steady state $n_0=\{ \langle
b^{\dag }b\rangle _{q}\} _{cl}\left( t\rightarrow \infty \right) $.
From the linearized Hamiltonian \eqref{Eq:DispHam} we can derive a
set of coupled differential equations for the quantum averages, e.g.
$\langle b^\dag b\rangle_q$, which can be solved in the perturbative
limit $g_{0}|\alpha(t)|\ll \kappa $. Averaging the result over the
classical noise process~\cite{Average} we obtain an effective
cooling equation
\begin{equation}  \label{eq:cooling}
\dot n=-W(n-n_{0})\,,
\end{equation}%
with $n=\{\langle b^{\dag }b\rangle _{q}\}_{cl}$. Here the cooling rate $%
W=S(\omega _{m})-S(-\omega _{m})$ and the average steady state
occupation
number $n_{0}=S(-\omega _{m})/W$ depend on the total noise spectrum $%
S(\omega )=S_{N}(\omega )+S_{A}(\omega )$ which is the sum of the
modulated intracavity amplitude correlation spectrum
\begin{equation}  \label{eq:SA}
S_{A}(\omega )=2g_{0}^{2}\mathrm{Re}\int_{0}^{\infty }d\tau
\,\left\{ \alpha ^{\ast }(\tau )\alpha (0)\right\}
_{cl}e^{i\Delta\tau }\,e^{-\kappa \tau }e^{i\omega \tau }\,,
\end{equation}
and the spectrum of intensity fluctuations,
\begin{equation}  \label{eq:SN}
S_{N}(\omega )=2g_{0}^{2}\mathrm{Re}\int_{0}^{\infty }d\tau
\,\left\{ \mathcal{N}(\tau )\mathcal{N}(0)\right\} _{cl}\,e^{i\omega
\tau }.
\end{equation}%
For the case of a non-fluctuating driving field $\mathcal{E}(t)=\mathcal{E}%
_{0}$, $\alpha(\tau)= \alpha_{0}=\mathcal{E}_{0}/(i\Delta+\kappa)$ and $S_N(\omega)=0$. Then, by evaluating Eq.~\eqref{eq:SA} for the optimized detuning $\Delta=\Delta_{op} =-%
\sqrt{\kappa^2+\omega^2_{m}}$  we recover the standard results,
\begin{equation}  \label{Eq:SideCool}
W_0= \frac{2g_{0}^{2}|\alpha _{0}|^{2}}{\kappa }\left(\frac{\omega_m}{%
|\Delta_{op}|}\right),\qquad n_0=\frac{1}{2}\left(\frac{|\Delta_{op}|}{%
\omega_m}-1\right),
\end{equation}%
which are well known from the theory of opto-mechanical laser cooling \cite%
{WislonRaePRL2007,MarquardtPRL2007}. In particular in the sideband
resolved
regime $\omega_m\gg \kappa$ the minimal occupation number is $%
n_0=\kappa^2/4\omega_m^2$.

Our goal in the following is to study the
influence of a noisy driving field on the final temperature. From the definition of $\alpha(t)$ given in Eq.~\eqref{Eq:alpha} and $%
S_A(\omega)$, $S_N(\omega)$ given in Eqs.~\eqref{eq:SA}
and~\eqref{eq:SN} we see that the final occupation number $n_0$
depends on integrals over two and four point correlation functions
of the stochastic quantity $e^{i\phi(t)}$, which itself is a
non-linear function of the noise process $\dot \phi(t)$.  For the
remainder of the paper we therefore assume that the total change in
phase $\phi(t)$ accumulated on a timescale of $\kappa$ is small.
This assumption is equivalent to the experimentally relevant limit
$\Gamma_L\ll \kappa$ and allows a rigorous expansion of noise
correlation functions in powers $\Gamma_L/ \kappa$.
 The details of these calculations are shifted to
App.~\ref{app:A} and~\ref{app:B} and here we summarize only the main
results.

\subsection*{Amplitude fluctuations}

Let us first look at phase noise induced modifications of
$S_A(\omega)$. To provide some physical insight we can assume in a
zeroth order approximation that  the cavity field simply follows the
driving field adiabatically,
\begin{equation}\label{eq:alphaApp}
\alpha(t)\approx \mathcal{E}_{0}e^{i\phi (t)}/(\kappa-i\Delta).
\end{equation}
By inserting this expression into the definition of $S_A(\omega)$ we
obtain
\begin{equation}  \label{eq:SAapprox}
S_{A}(\omega)/W_0 \simeq \kappa \mathrm{Re}\int_{0}^{\infty }d\tau
\,\{ e^{i\phi(\tau )}e^{-i\phi(0)}\} _{cl}\, e^{i(\Delta+\omega)\tau
}\,e^{-\kappa \tau }\,,
\end{equation}
and a simple result for this integral can be found in the white
noise limit $\gamma
_{c}\rightarrow \infty $ where 
$\{ e^{i\phi(\tau )}e^{-i\phi(0)}\} _{cl}\simeq e^{-\Gamma_L |\tau|} $. %
From this estimate we expect correction to $S_A(\pm \omega_m)$ and
therefore to the final occupation number $n_0$ which are of order
$\mathcal{O}(\Gamma_L/\kappa)$. 


In App.~\ref{app:A}  we present a more rigorous calculation of
$S_A(\omega)$ for $\Gamma_L/\kappa\ll 1$ and here briefly discuss
results for the resolved sideband regime $\kappa\ll\omega_m$.  For
transitions on the red sideband, which are associated with a resonant exchange of a vibrational quanta and a cavity photon, we obtain
\begin{equation}
\frac{S_A(\omega_m)}{W_0}\simeq 1 - \int \frac{d\Omega}{2\pi} \,
\frac{ S_{\dot \phi}(\Omega)}{\kappa^2+\Omega^2} \geq 1-
\frac{\Gamma_L}{\kappa}.
\end{equation}
Here we obtain a dependence on the low frequency part of the noise spectrum and the lower bound which has been derived using the model defined in
Eq.~\eqref{eq:Noise} agrees with the estimates from above. For the blue sideband transitions, which correspond to a simultaneous excitations of a photon and a phonon we obtain instead
\begin{equation}\label{eq:BlueSideband}
\frac{S_A(-\omega_m)}{W_0}\simeq \frac{\kappa^2}{4\omega_m^2} +
\frac{\kappa}{18\omega_m^2}\left[S_{\dot
\phi}(\omega_m)+\frac{1}{4}S_{\dot \phi}(2\omega_m)\right],
\end{equation}
i.e., a modification which depends on phase noise at high
frequencies. The contribution at $2\omega_m$ can be understood form
the fact that for $\Delta=-\omega_m$ it takes this amount of energy to excite a photon and
a phonon. Although noise at $\omega_m$ is
non-resonant with this excitation process it is maximally enhanced
by the cavity response, and therefore leads to an equivalent
contribution. However, in both cases the noise is either
non-resonately driving the system or is suppressed by the cavity
response and the resulting corrections to the final occupation
number only scale as $\sim \kappa^2/\omega_m^2\times
\Gamma_L/\kappa$, i.e. they are reduced by the sideband parameter
$\kappa/\omega_m$. In summary we conclude that phase modulations of
the cavity field lead to a reduction of the cooling rate by
$\Gamma_L/\kappa$, the ratio of laser to cavity linewidth, and
therefore pose no serious experimental limitation.

%

\subsection*{Intensity fluctuations}

We now look at intensity fluctuations. As the zeroth order approximation of $%
\alpha(t)$ would lead to a constant intensity $|\alpha(t)|^2$ we
expect the first correction for a slow noise process to scale as
$\sim \dot \phi(t)/\kappa$ and we therefore approximate
\begin{equation}
\alpha(t)\approx \frac{\mathcal{E}_{0}e^{i\phi (t)}}{\kappa-i\Delta}%
\,\left(1-i \frac{\dot{\phi} (t)}{\kappa-i\Delta }\right).
\end{equation}
Inserting this expression into the definition of $S_N(\omega)$ in Eq.~%
\eqref{eq:SN} we obtain for $\Delta=-\omega_m$,
\begin{equation}  \label{eq:SNapprox}
\frac{S_N(\omega_m)}{W_0}\sim |\alpha_0|^2 \mathrm{Re} \int_0^\infty
\{ \dot \phi(\tau)\dot \phi(0)\}_{cl} \, e^{i\omega_m \tau} d\tau .
\end{equation}
Already form this simple estimate we find that in accordance with
earlier predictions~\cite{Schliesser2008,DiosiPRA2008} intensity
fluctuations do impose a limit on the final occupation number which
increases with the mean cavity photon number $|\alpha_0|^2$. However, Eq.~%
\eqref{eq:SNapprox} also predicts a crucial dependence of this limit
on the particularities of the noise process, i. e., the correlation
function $\{ \dot \phi(\tau)\dot \phi(0)\}_{cl}$. This means that
the final occupation number depends in a non-universal way on the
phase noise characteristics and cannot be inferred from a white
noise model described  by the linewidth only.

The basic prediction of Eq.~\eqref{eq:SNapprox} is confirmed by a
more rigorous derivation outlined in App.~\ref{app:B} where we
evaluate the correlation function $\left\{ \mathcal{N}(\tau
)\mathcal{N}(0)\right\} _{cl}$ to first order in the parameter
$\Gamma_L/\kappa$. Under this assumption and $\Delta=\Delta_{op}$
the resulting limit on the final occupation number can be written as
\begin{equation}  \label{eq:Limit}
n_0\geq \frac{S_N(\omega_m)}{W}\simeq |\alpha_0|^2 \frac{ S_{\dot
\phi}(\omega_m)}{2\kappa}
\left(\frac{|\Delta_{op}|}{\omega_m}\right).
\end{equation}
This is the main result on laser noise induced limitations of
opto-mechanical cooling
in the weak coupling regime. It agrees with a simplified analysis given in Ref.~\cite{Schliesser2008} for the resolved sideband regime and generalizes the result for the white noise limit derived in Ref.~\cite{DiosiPRA2008} for arbitrary noise processes. It is instructive to consider the toy model for laser noise as given in Eq.~%
\eqref{eq:Noise} from which we obtain in the sideband resolved
regime
\begin{equation}
n_0\geq |\alpha_0|^2 \frac{\Gamma_L}{\kappa}\frac{\gamma_c^2}{%
\gamma_c^2+\omega_m^2}.
\end{equation}
For typical experimental parameters $|\alpha_0|^2 \approx 10^{10}$, $%
\omega_m\approx 10$ MHz, $\kappa \approx 1$ MHz and $\Gamma_L
\approx 1$ kHz the assumption of a white noise model,
$\gamma_c\rightarrow \infty$, would lead to the prediction $n_0
\gtrsim 10^5$, which is in sharp contrast with observed experimental
data. However, for a realistic noise model with a finite cutoff
frequency $\gamma_c \ll \omega_m$ intensity fluctuations are
strongly suppressed and the resulting limits $n_0\geq 1-100$ are
consistent with recent experiments.

It was pointed out already in \cite{Schliesser2008} that the scaling
of the bound in \eqref{eq:Limit} with the intracavity photon number
$|\alpha_0|^2$ implies an optimal driving power balancing the
cooling effects with heating due to laser noise. If we combine the
present result with the known bounds from the theory of
opto-mechanical cooling with an ideal laser
\cite{MarquardtPRL2007,WislonRaePRL2007} we get
\begin{equation}
  n_0\simeq \frac{2\kappa\Gamma_m}{g_0^2|\alpha_0|^2}+\frac{\kappa^2}{4\omega_m^2}+\frac{|\alpha_0|^2}{2\kappa}S_{\dot
\phi}(\omega_m).
\end{equation}
Here $\Gamma_m=k_BT/\hbar Q_m$ is the mechanical heating rate for a
resonator with quality factor $Q_m$ coupled to a thermal phonon
reservoir of temperature $T$. The first term on the right hand side
is the residual thermal occupation, the middle term is the
contribution from heating due to Stokes scattering, and the last
term stems form laser noise. For an optimal choice of the
intracavity photon number
$|\alpha_0|^2=(2\kappa)^2\Gamma_m/g_0^2S_{\dot\phi}(\omega_m)$ we
obtain
\begin{equation}
n_0\simeq 2 \sqrt{\frac{\Gamma_m S_{\dot
\phi}(\omega_m)}{g_0^2}}+\frac{\kappa^2}{4\omega_m^2}.
\end{equation}
From this result we identify Eq.~\eqref{eq:Cond1} 
 as the relevant condition to achieve ground state cooling in weakly
coupled opto-mechanical systems, provided the optimal value of
$|\alpha_0|^2$ is not prohibitive. Note that this condition depends
on $g_0$, the radiation pressure coupling per \textit{single
photon}, cf. Eq.~\eqref{eq:ham}.

The details of the laser noise characteristics will depend on the
concrete experimental setup, but the spectrum of intensity
fluctuations $S_N(\omega)$ can in each case be directly measured,
e.g., from correlations of the transmitted intensity
$I_{\mathrm{out}}(t)$. In the strongly driven regime
the cavity out-field is dominated by the classical part $b_{\mathrm{out}%
}(t)\sim \sqrt{\kappa}\alpha(t)$ and using standard results on
photon counting statistics~\cite{QuantumNoise} we obtain the simple
relation
\begin{equation}
\frac{S_N(\omega)}{W_0 }= \frac{1}{2}\left[\frac{S_I(\omega)}{S_{sn}}-1%
\right]\,.
\end{equation}
Here $S_I(\omega)= \int_{-\infty}^\infty d\tau \,C_I(\tau)
e^{i\omega \tau}$
is the spectrum of the normalized photon current correlation function $%
C_I(\tau)=\{I_{\mathrm{out}}(\tau)I_{\mathrm{out}}(0)\}_{cl}/\{I_{\mathrm{out%
}}(0)\}^2_{cl}-1$ and $S_{sn}$ is the shot noise contribution
thereof.

\section{Strong coupling regime}\label{sec:strong}

We now consider an opto-mechanical system operated in the strong
coupling regime where the linear photon-phonon interaction strength
$G=g_{0}\alpha _{0}$ exceeds the cavity linewidth $\kappa $. This
regime has been recently studied experimentally
\cite{Groblacher2009} and has been discussed
theoretically in the context of opto-mechanical cooling \cite%
{WilsonRaeNJP2008,Dobrindt2008}. However, more importantly strong
coupling conditions enable a coherent exchange of photonic and
mechanical excitations. Thus, the opto-mechanical system can serve
as a quantum interface between photons and phonons with potential
applications for state preparation and quantum measurements of
macroscopic mechanical motion. It is therefore worthwhile to study
the role of phase noise in particular under strong coupling
conditions where due to the required large values of $|\alpha _{0}|$
more pronounced effects are expected.

\subsection*{Strong coupling}

Let us first briefly review the main features of the opto-mechanical
system in the strong coupling regime, ignoring for the moment the
presence of phase noise or other imperfections. When the interaction
between the cavity and the resonator mode exceeds the cavity decay
rate the system dynamics is conveniently described in terms of the
new collective operators $A_{\pm }$ which diagonalize
$H_{\mathrm{opt}}$,
\begin{equation}
H_{\mathrm{opt}}=\omega _{+}A_{+}^{\dag }A_{+}+\omega
_{-}A_{-}^{\dag }A_{-}\,.
\end{equation}%
Assuming resonance conditions $\Delta _{c}=-\omega _{m}$ and
$|G|<\omega _{m}/2 $ which is required for stability
\cite{GenesPRA2008}, the eigenfrequencies are $\omega _{\pm }=\omega
_{m}(1\pm 2|G|/\omega _{m})^{1/2}$ and the normal
modes are approximately given by 
\begin{equation}\label{eq:Apm}
A_{\pm }\simeq \frac{e^{-i\theta }}{\sqrt{2}}\,b\pm \frac{e^{i\theta }}{%
\sqrt{2}}\,a\mp \frac{G^*}{2\omega _{m}}\left( \frac{e^{i\theta }}{\sqrt{2}}%
\,b^{\dag }\pm \frac{e^{-i\theta }}{\sqrt{2}}\,a^{\dag }\right) ,
\end{equation}%
where we have defined $e^{i2\theta }:=\alpha _{0}/|\alpha _{0}|$.
For not too large values of $|G|$ the eigenmodes are essentially
equal superpositions of the original cavity and resonator mode and
they are split in frequency by $\omega _{+}-\omega _{-}\simeq 2|G|$.
A normal mode splitting exceeding the cavity decay rate $\kappa $ is
a first signature of
the strong coupling regime and has recently been observed in experiments \cite%
{Groblacher2009}. In this limit the Liouville operator is
approximately given by \cite{WilsonRaeNJP2008}
\begin{equation}\label{eq:MEStrong}
\begin{split}
\mathcal{L}_{f}(\rho )\simeq & \frac{\kappa }{2}(n_{0}+1)\sum_{\xi
=\pm }\left( 2A_{\xi }\rho A_{\xi }^{\dag }-A_{\xi }^{\dag }A_{\xi
}\rho -\rho
A_{\xi }^{\dag }A_{\xi }\right)  \\
+& \frac{\kappa }{2}n_{0}\sum_{\xi =\pm }\left( 2A_{\xi }^{\dag
}\rho A_{\xi }-A_{\xi }A_{\xi }^{\dag }\rho -\rho A_{\xi }A_{\xi
}^{\dag }\right).
\end{split}%
\end{equation}%
As expected we see that both modes decay with half of the cavity
decay rate. A lower limit on the achievable occupation numbers
$n_{0}=|G|^{2}/4\omega
_{m}^{2}\ll 1$ arises from small admixtures of $%
a^{\dag }$ and $b^{\dag }$ in Eq.~\eqref{eq:Apm} due to energy non-conserving terms in $%
H_{opt}$. Neglecting this small correction we immediately see from Eq.~\eqref{eq:MEStrong} that without any additional
heating mechanisms the total number of excitations in the system
$n_{\mathrm{tot}}=\langle A^\dag_+A_+\rangle+\langle
A^\dag_-A_-\rangle$ decays as
\begin{equation}
\dot n_{\mathrm{tot}} = - \kappa  n_{\mathrm{tot}}.
\end{equation}
Therefore, in the strong coupling limit the opto-mechanical cooling rate $W $ is independent of the driving strength and
saturates at the maximum value set by the cavity field decay rate
$\kappa$.

\subsection*{Strong coupling in the presence of phase noise}

In the presence of noise the picture above is modified on one hand
by the presence of fluctuating forces and on the other hand by a
modulated cavity-resonator coupling $G\rightarrow G(t):=g_{0}\alpha
(t)$. To account
for this time-dependent coupling we introduce time-dependent mode operators $%
A_{\pm }(t)$ which diagonalize the Hamiltonian $H_{\mathrm{opt}}(t)$
given in Eq.~\eqref{Eq:DispHam}
 at each point in time, i.e.
\begin{equation}
\lbrack H_{\mathrm{opt}}(t),A_{\pm }(t)]=-\omega _{\pm }(t)A_{\pm
}(t).
\end{equation}%
Here $\omega _{\pm }(t)=\omega _{m}(1\pm 2|G(t)|/\omega _{m})^{1/2}$
are the instantaneous eigenfrequencies and the decomposition of
$A_{\pm }(t)$ contains now also time dependent phases $e^{i2\theta
(t)}:=\alpha (t)/|\alpha (t)|$. To simplify the following discussion
and to focus on the effect of noise we assume that the condition
$\kappa \ll |G(t)|\ll \omega _{m}$ is strictly fulfilled, which
allows us to neglect in Eq.~\eqref{eq:Apm} contributions of order
$\mathcal{O}(G(t)/\omega _{m})$. Under this assumption
\begin{equation}
\dot{\rho}=-i[H(t),\rho ]+\mathcal{L}_{f}(t)(\rho ),
\end{equation}%
where
\begin{equation}
H(t)\!=\!\sum_{\xi =\pm }\omega _{\xi }(t)A_{\xi }^{\dag }(t)A_{\xi }(t)+%
\frac{g_{0}}{\sqrt{2}}\left( e^{i\theta (t)}A_{\xi
}(t)\!+\!e^{-i\theta (t)}A_{\xi }^{\dag }(t)\right) \mathcal{N}(t),
\end{equation}%
and
\begin{equation}
\mathcal{L}_{f}(t)(\rho )\simeq \frac{\kappa }{2}\sum_{\xi =\pm
}\left( 2A_{\xi }(t)\rho A_{\xi }^{\dag }(t)-A_{\xi }^{\dag
}(t)A_{\xi }(t)\rho -\rho A_{\xi }^{\dag }(t)A_{\xi }(t)\right) .
\end{equation}%
We now investigate how those modifications affect cooling and
coherent dynamics in the strong coupling regime.


\subsection*{Cooling}

We first look at phase noise induced limitations for ground state
cooling in the strong coupling regime. In the previous discussion on
the weak coupling regime we have seen that the effect of the
modulation of $\alpha (t)$ is less crucial for cooling than that of
the fluctuations of the cavity intensity. We therefore neglect for
the moment the explicit time dependence of $A_{\pm }(t)$ and study
the effect of $\mathcal{N}(t)$ only. A justification for this approximation follows form the analysis present below. We obtain the coupled equations
\begin{eqnarray}
\langle \dot{N}_{\pm }\rangle  &=&-\kappa \langle N_{\pm }\rangle -ig_{0}%
\mathcal{N}(t)\langle e^{-i\theta }A_{\pm }^{\dag }-e^{i\theta
}A_{\pm
}\rangle , \\
\langle \dot{A}_{\pm }\rangle  &=&-(i\omega _{\pm }+\kappa
/2)\langle A_{\pm }\rangle -ig_{0}\mathcal{N}(t)e^{-i\theta },
\end{eqnarray}%
where $N_{\pm }=A_{\pm }^{\dag }A_{\pm }$. After averaging over the
noise we end up with
\begin{equation}
\{\langle \dot{N}_{\pm }\rangle \}_{cl}=-\kappa \{\langle N_{\pm
}\rangle \}_{cl}+S_{N}(\omega _{\pm }+i\kappa /2),
\end{equation}%
where $S_{N}(\omega )$ is defined in Eq.~\eqref{eq:SN}. In the limit
$\Gamma _{L}/\kappa \ll 1$ we can use the results derived in
App.~\ref{app:B} to evaluate this quantity. By assuming that for
frequencies $\Omega \sim \omega _{m}$ the noise spectrum
$S_{\dot{\phi}}(\Omega )$ is flat on a scale $\kappa $ and to lowest
order in $\kappa /G$ we obtain
\begin{equation}\label{eq:SNstrong}
S_{N}(\omega _{\pm }+i\kappa /2)\simeq g_{0}^{2}|\alpha
_{0}|^{4}\left[ \frac{4\omega _{m}^{2}S_{\dot{\phi}}(\omega _{\pm
})}{(\omega _{\pm }^{2}-\omega
_{m}^{2})^{2}}+\frac{S_{\dot{\phi}}(\omega _{m})}{2(\omega _{\pm
}-\omega _{m})^{2}}\right] .
\end{equation}%
Interestingly, heating arises from two contributions. As expected,
the first
term represents intensity fluctuations at the eigenfrequency $\omega _{\pm }$%
. Since in the strong coupling regime these frequencies are well
separated
from the cavity resonance, the noise at these frequencies is suppressed by $%
\omega _{m}^{2}/(\omega _{\pm }^{2}-\omega _{m}^{2})^{2}$. The
second contribution in Eq.~\eqref{eq:SNstrong} represents noise
which is resonantly enhanced by the cavity. Although it is not
resonant with $\omega _{\pm }$ it still couples to the damped motion
of $A_{\pm }$
and we see that both terms lead to similar contributions for
heating.  For $\omega _{\pm }\simeq \omega _{m}\pm G$ the lower
limit for the total steady state occupation number is then given by
\begin{equation}
n_{\mathrm{tot}}\geq |\alpha
_{0}|^{2}\frac{S_{\dot{\phi}}(\omega _{m})+S_{\dot{\phi}}(\omega
_{+})+S_{\dot{\phi}}(\omega _{-})}{\kappa }.
\end{equation}%
Surprisingly for $S_{\dot{\phi}}(\omega _{m})\approx
S_{\dot{\phi}}(\omega _\pm)$ this result is quite similar to the
weak coupling regime which we attribute to a cancellation of two
effects. On one hand for increasing $\alpha _{0}$ the cooling rate
saturates in the strong coupling regime at $W\simeq \kappa $. One
would therefore naively expect a scaling $n_{0}\sim |\alpha
_{0}|^{4}$. However, since the eigenfrequencies $\omega _{\pm }$ are
well detuned from
the cavity resonance the effect of phase fluctuations is suppressed by $%
1/G^{2}\sim 1/g_{0}^{2}|\alpha _{0}|^{2}$ (see
Eq.~\eqref{eq:SNstrong}) which reduces in the strong coupling limit
the scaling from $|\alpha _{0}|^{4}$ to $|\alpha _{0}|^{2}$.

Due to the saturation of the cooling rate the strong coupling regime
does not offer a particular advantages for cooling. We therefore do
not go further into details and study instead the effects of phase
noise for coherent dynamics where strong coupling conditions are
essential.

\subsection*{Coherent oscillations}

As already mentioned above, the interesting aspect about the strong
coupling regime of opto-mechanical systems is the ability to realize
a coherent interface between mechanical and optical modes. For
example, if  the resonator is initially prepared in the ground state
(possibly in combination with other cooling methods) and the cavity
in a Fock state $|1\rangle_c$ this single excitation is swapped onto the
resonator mode at time $t_s=\pi/(2|G|)$ and recovered at a later
time $t_r=\pi/|G|$. More general we can describe this process in terms of an arbitrary 
coherent state $|\xi_c\rangle_c$ which in a frame rotating with $\omega_r$ evolves
under ideal conditions as
\begin{equation}\label{eq:StateTransfer}
|0\rangle_r|\xi_c\rangle_c \rightarrow |-i e^{-i2\theta}
\xi_c\rangle_r|0\rangle_c \rightarrow|0\rangle_r|-\xi_c\rangle_c.
\end{equation}
Since this evolution is independent of the initial coherent state
amplitude it can be generalized to arbitrary quantum states.
Therefore, apart from a known phase the dynamics generated by $H_{op}$ implements a faithful mapping between the states of the cavity and the resonator mode.  If this operation is fast compared to photon loss,
i.e. $\kappa\ll |G|$, it can be employed as a coherent way for an
optical preparation of motional states. Alternatively the reverse
process would enable an optical detection of the resonator state.

To characterize coherent oscillations in the presence of photon loss
and phase noise it is sufficient to study the evolution of an
initial coherent state $|\psi_0\rangle=
|\xi_r(0)\rangle_r|\xi_c(0)\rangle_c$ under the evolution of the
effective Hamiltonian $H_{\rm eff}(t)=H(t)-i\kappa c^\dag c$.
Ignoring small corrections of order $G/\omega _{m}\ll 1$ which are
not essential in the following discussion the state will then evolve
into $|\psi(t)\rangle= |\xi_r(t)\rangle_r|\xi_c(t)\rangle_c$ where
the corresponding coherent state amplitudes can be written as
\begin{equation}
\left(\begin{matrix}
\xi_r(t)  \\
\xi_c(t)
\end{matrix}
\right)= \left(\begin{matrix}
c_{bb}(t) &  c_{ba}(t)  \\
c_{ab}(t) & c_{aa}(t)
\end{matrix}\right)
\left(\begin{matrix}
\xi_r(0)  \\
\xi_c(0)
\end{matrix}\right) +
\left(\begin{matrix}
c_b(t)  \\
c_a(t)
\end{matrix}\right).
\end{equation}
The ideal evolution described in Eq.~\eqref{eq:StateTransfer}
suggest to use the state overlaps $|\langle
-i\xi_c(0)e^{-i2\theta}|\xi_c(t)\rangle|^2$ and  $|\langle
-\xi_c(0)|\xi_c(t)\rangle|^2$ to characterize a state transfer or a
full oscillation respectively. We here choose the latter option and
for an distribution $P(r,\theta)$ of initial coherent state
amplitudes $\xi_c(0)=re^{i\phi}$  we define the fidelity
\begin{equation}
\mathcal{F}(t)=  \frac{1}{\pi}\int_0^\infty r dr  \int_0^{2\pi}
d\phi \,\{ e^{-|\xi_c(t)+r e^{i\phi}|^2}\}_{cl} P(r,\theta).
\end{equation}
To be more concrete we average over initial states with $r=1$ and
for $\xi_r(0)=0$ we finally end up with
\begin{equation}\label{eq:Fidelity}
\mathcal{F}(t)=e^{-\{|c_{aa}(t)+1|^2\}_{cl}} \times
e^{-\{|c_{a}(t)|^{2}\}_{cl}}.
\end{equation}
While there is some arbitrariness in this choice of a fidelity, the definition~\eqref{eq:Fidelity} is sensitive to different aspects of the noise and should therefore be a good characterization for coherent processes involving a low number of excitations.
For a noiseless system $c_a(t)=0$ and we obtain
\begin{equation}
\mathcal{F}_{0}(t)= e^{- |1+\cos(|G|t)e^{-\kappa t/2}|^2}.
\end{equation}%
The fidelity for a full oscillation is approximately given by $%
\mathcal{F}_0(t_r)\simeq e^{-\pi^2 \kappa^2 /|G|^2}$.


To study the additional degrading of $\mathcal{F}(t)$ in the
presence of noise we write $a(t)\simeq
e^{-i\theta(t)}(A_{+}(t)-A_{-}(t))/\sqrt{2}$. The coefficients
$c_{aa}(t)$ and $c_{a}(t)$ can then be calculated from the evolution
of the mode operators,
\begin{equation}\label{eq:ApmEvo}
i\left(
\begin{matrix}
 \dot{A}_{+}  \\
 \dot{A}_{-}
\end{matrix}%
\right) =\left(
\begin{matrix}
\omega _{+}(t)\!-\!i\kappa /2 & -\dot{\theta}(t) \\
-\dot{\theta}(t) & \omega _{-}(t)\!-\!i\kappa /2%
\end{matrix}%
\right) \left(
\begin{matrix}
 A_{+} \\
 A_{-}
\end{matrix}%
\right) +g_{0}\mathcal{N}(t)\left(
\begin{matrix}
e^{-i\theta (t)} \\
e^{-i\theta (t)}%
\end{matrix}%
\right) .
\end{equation}%
From this expression we identify three potential effects of phase noise. First, intensity fluctuations $\sim \mathcal{N}(t)$ introduce a random displacement $%
c_{a}(t)$ which is related to the additional heating discussed
above. Second, the fluctuating cavity amplitude $|\alpha (t)|$ leads
to a fluctuating normal mode splitting, $\omega _{\pm }(t)\simeq
\omega _{\pm }\pm \delta \omega (t)$ where  $\delta \omega (t)\simeq
g_{0}\mathcal{N}(t)/|\alpha _{0}| $. Third, phase fluctuations of
the intra-cavity field  causes non-adiabatic transitions between the
modes $A_{\pm }$ proportional to $\dot{\theta}(t)$.

\subsection*{High frequency noise}
Based on our discussions so far we expect that intensity fluctuations
are the dominant decoherence mechanism. In a first approximation we therefore
neglect the time dependence of  $\omega(t)$ and $\theta(t)$ and
study the effect of $\mathcal{N}(t)$ only. By integrating
Eq.~\eqref{eq:ApmEvo} we then obtain
\begin{equation}
\mathcal{F}(t)=\mathcal{F}_0 \times e^{-D(t)},
\end{equation}
where
\begin{equation}
\begin{split}
D(t)=\{|c_{a}(t)|^{2}\}_{cl}=& 2g_{0}^{2}\int_{0}^{t}d\tau
\int_{0}^{t}d\tau ^{\prime }e^{i\omega _{m}(\tau -\tau ^{\prime
})}e^{-\kappa (\tau +\tau
^{\prime })/2} \\
\times & \sin (|G|\tau )\sin (|G|\tau ^{\prime })\{\mathcal{N}(\tau )%
\mathcal{N}(\tau ^{\prime })\}_{cl}.
\end{split}%
\end{equation}%
For a full oscillation period $t=\pi /(|G|)$ and $|G|\gg \kappa $
this expression reduces to
\begin{equation}
D(t=\pi /(|G|))\approx 8|\alpha _{0}|^{2}\frac{S_{\dot{\phi}}(\omega _{m})}{%
\kappa }.
\end{equation}%
As a consequence the error $\epsilon =1-\mathcal{F}$ for a full
oscillation between cavity and resonator mode is
\begin{equation}
\epsilon \approx \frac{\pi \kappa }{g_{0}|\alpha _{0}|}+8|\alpha
_{0}|^{2}\frac{S_{\dot{\phi}}(\omega _{m})}{\kappa }.
\end{equation}%
For fixed cavity parameters there is an optimal field amplitude
$|\alpha _{0}|$ for which, apart from a numerical prefactor, the
error scales as
\begin{equation}
\epsilon \approx \sqrt[3]{\kappa S_{\dot{\phi}}(\omega
_{m})/g_{0}^{2}}.
\end{equation}%
We see that apart from other imperfections, achieving $S_{\dot{
\phi}}(\omega _{m})\ll g_{0}^{2}/\kappa$ is a necessary requirement
for the observation of coherent dynamics in opto-mechanical systems.
Since $\kappa$ must also exceed the mechanical heating rate
$\Gamma_m$ this result implies that the observation of coherent
oscillations puts more stringent bounds on the acceptable level of
phase noise than just cooling.

\subsection*{Low frequency noise}
While due to the scaling $\sim |\alpha_0|^2$ intensity fluctuations
impose a sever limitation on coherent oscillations, this effect
depends on phase noise at relatively high frequencies $\omega \sim
\omega _{m}$. Since the relevant system dynamics occurs on a slower
timescale $|G|^{-1}$  this noise can in principle be filtered
out in a carefully designed experimental setup. An important
question therefore is whether or not other decoherence mechanisms
exist which depend on low frequency regime of the phase noise
spectrum.

To address this question we now assume that the phase noise spectrum
$S_{\dot{\phi}}(\Omega )$ has relevant contributions only at $\Omega
\ll \omega _{m}$. This assumption allows us to
omit the term $\sim \mathcal{N}(t)$ in Eq.~\eqref{eq:ApmEvo} and to study the effects of $%
\delta \omega (t)$ and $\dot{\theta}(t)$ only.  For a simplified discussion we will also neglect cross-correlations between these two stochastic quantities. In App.~\ref{app:C} we show  that the resulting fidelity is then given by
\begin{equation}\label{eq:FidelityLowFreq}
\begin{split}
\mathcal{F}(t)\approx \exp\left(- \{|1+\cos(|G|t)e^{-\kappa t/2}e^{-W(t)/2}e^{-R(t)/2}|^2\}_{cl}\right).
\end{split}
\end{equation}
Compared to the case of a noiseless laser we obtain two additional contributions to decoherence. Here 
\begin{equation}\label{eq:W}
W(t)=2\int_{0}^{t}ds\int_{0}^{s}ds^{\prime }\,\{\delta \omega
(s)\delta \omega (s^{\prime })\}_{cl},
\end{equation}%
describes dephasing due to a modulated normal mode splitting and
\begin{equation}\label{eq:R}
R(t)=2{\rm Re}\int_{0}^{t}ds\int_{0}^{s}ds^{\prime }\,e^{iG(s-s^{\prime })}\{\dot{%
\theta}(s)\dot{\theta}(s^{\prime })\}_{cl},
\end{equation}%
describes non-adiabatic transitions between the normal modes.

As anticipated above Eq.~\eqref{eq:W} and Eq.~\eqref{eq:R} differ
qualitatively form the heating discussed above in the sense that the
quantities $W(t)$ and $R(t)$ depend on noise far
below the resonator frequency. To see this more explicitly we use $\delta \omega (t)=g_{0}%
\mathcal{N}(t)/|\alpha _{0}|$ and obtain
\begin{equation}
W(t)\approx \frac{g_{0}^{2}|\alpha _{0}|^{2}}{\omega
_{m}^{2}}\int_{|\Omega |<\omega _{m}}\frac{d\Omega }{2\pi
}\,\frac{4\sin ^{2}(\Omega t/2)}{\Omega ^{2}}\,S_{\dot{\phi}}(\Omega
).
\end{equation}%
An upper bound for this integral can be obtained form the long time
limit,
\begin{equation}
W(t)\leq \frac{|G|^{2}}{\omega _{m}^{2}}S_{\dot{\phi}}(0)\times t.
\end{equation}%
We see a dependence on zero frequency noise $S_{\dot{\phi}%
}(0)\approx \Gamma _{L}$, but since this noise is far detuned from
the cavity resonance this dephasing process is suppressed by
$|G|^{2}/\omega _{m}^{2}$. For the parameter regime of interest the
resulting error for a coherent oscillation  $\epsilon \approx
|G|\Gamma _{L}/\omega _{m}^{2}\ll 1$ is therefore always smaller
than the error due to photon loss.

A similar conclusion can be obtained for decoherence caused by
non-adiabatic transitions. A simple estimate for the upper bound on
$R(t)$ can be obtain by
assuming that the intracavity phase $2\theta (t)$ follows the laser phase $%
\phi (t)$ adiabatically (see Eq.~\eqref{eq:alphaApp}). Under this approximation we obtain
\begin{equation}
R(t)\leq \frac{1}{4}S_{\dot{\phi}}(|G|)\times t.
\end{equation}%
Again we see that for $\Gamma _{L}\ll \kappa $ this decoherence
process is negligible compared to cavity loss and a more accurate
calculation predicts a further reduction due to a suppression
of low frequency noise by the cavity response. We conclude that
while low frequency noise does affect the dynamics of the
opto-mechanical system it is for the parameter regime of interest
negligible compared to photon loss.

\section{Summary $\&$ conclusions}

In summary we have analyzed the effect of laser phase noise on
cooling and coherent dynamics in a generic opto-mechanical system
comprised of a driven optical cavity with a vibrating end-mirror.
Our approach significantly extends and generalizes previous treatments to this problem and provides a rigorous way to study different aspect of the laser noise in such systems.  For opto-mechanical cooling in the weak coupling regime our predictions 
our predictions for a final occupation number $n_{0}\sim \mathcal{O}(10)$ 
are consistent with experimental data and we derive a condition for the noise spectrum which is required in future experiments to achieve ground state cooling. We have also show that the discrepancy between experiments and previous theoretical calculations \cite{DiosiPRA2008} is based on the assumption of a white noise model, which does not lead to physically meaningful predictions. For opto-mechanical systems in the strong coupling regime  we have shown that the effects of phase noise still scale linearly with the intensity but the observation of coherent oscillations places more stringent bounds on tolerable level of phase noise. Nevertheless, we conclude that ground state cooling and coherent state transfer experiments can be achieved with state of the art laser stabilization techniques.

\paragraph*{Acknowledgments}
The authors thank L. Diosi,  A. Schliesser, D. Vitali, J. Ye and P. Zoller for stimulating discussions. P. R. acknowledges support by the NSF through a grant for ITAMP. C.G. is thankful for support from Euroquam Austrian Science Fund project I1 19 N16 CMMC and K. H., C. G. and M.A. announce support by the Austrian Science Foundation under SFB FOQUS.

\begin{appendix}

\section{Evaluation of the amplitude fluctuation spectrum}\label{app:A}
We evaluate the amplitude fluctuation spectrum $S_A(\omega)$ defined
in Eq.~\eqref{eq:SA} in the limit $\Gamma_L/\kappa\ll 1$. From the
definition of $\alpha(t)$ given in Eq.~\eqref{Eq:alpha} we find that
under stationary conditions the correlation function of the
classical cavity field can be written in the form
\begin{equation}
\{ \alpha^*(\tau) \alpha(0)\}_{cl} =\int_{0}^\infty dy \,e^{-2\kappa
y} \int_{-2y}^{2y}dx \, e^{-i\Delta x}C_2(\tau -x )\,,
\end{equation}
where $C_2(t)=\{ \mathcal{E}^*(t)\mathcal{E}(0)\}_{cl}$ is the two
point correlation function of the driving field. For Gaussian phase
noise  we obtain
\begin{equation}\label{eq:C2}
C_2(t)= |\mathcal{E}_0|^2 e^{-  \frac{1}{2}\{ \phi^2(t)\}_{cl}},
\end{equation}
where $\phi(t)=\int_{-t/2}^{t/2}ds \,\dot \phi(s)$. Using the noise
model defined in Eq.~\eqref{eq:Noise} we see that the expectation
value in the exponent is bound by $\{ \phi^2(t)\}_{cl}\leq \Gamma_L
\gamma_c t^2$ for $t\lesssim \gamma_c^{-1}$ and  $\{
\phi^2(t)\}_{cl}\leq \Gamma_L t $  for long times. Therefore, in the
limit $\Gamma_L\ll \kappa$ we can expand the exponential in
Eq.~\eqref{eq:C2} to first order,
\begin{equation}
C_2(t)\simeq  |\mathcal{E}_0|^2\left(1-
\frac{1}{2}\int_{-t/2}^{t/2}ds \int_{-t/2}^{t/2}ds'\, \{\dot
\phi(s)\dot \phi(s')\}_{cl}\right).
\end{equation}
Equivalently, we can rewrite this expression in terms of the noise
spectrum,
\begin{equation}
C_2(t)\simeq |\mathcal{E}_0|^2 \left(1-    \int \frac{d\Omega}{2\pi}
S_{\dot \phi}(\Omega) \frac{1-\cos(\Omega t)}{\Omega^2}\right) .
\end{equation}
We insert this result back into the definition of $\{ \alpha^*(\tau)
\alpha(0)\}_{cl} $ and $S_A(\omega)$ and evaluate the remaining
integrals. Since the resulting general expressions are lengthy we
here only present the results for the sideband resolved regime
$\kappa \ll \omega_m$, where $\Delta_{op}=-\omega_m$ and
$W_0=2g_0^2|\alpha_0|^2/\kappa$. For the red sideband we obtain
\begin{equation}
\frac{S_A(\omega_m)}{W_0}\simeq 1 - \int \frac{d\Omega}{2\pi} \,
\frac{ S_{\dot \phi}(\Omega)}{\kappa^2+\Omega^2}.
\end{equation}
Using the noise model defined in Eq.~\eqref{eq:Noise} we see that in
the white noise limit $\gamma_c\gg \kappa$ corrections are of the
order of $ S_{\dot \phi}(0)/(2\kappa)\approx  \Gamma_L/\kappa$ while
in the opposite limit they scale as   $ \Gamma_L\gamma_c/\kappa^2
$. For the blue sideband transitions we obtain the result presented in Eq.~\eqref{eq:BlueSideband}.

\section{Evaluation of the intensity fluctuation spectrum}\label{app:B}

We evaluate the intensity fluctuation correlation function
$C_{\mathcal{N}}(\tau)=\{ \mathcal{N}(\tau) \mathcal{N}(0)\}_{cl}$
in the limit $\Gamma_L\ll \kappa$. Ignoring small corrections of
order $\mathcal{O}(g_0/\kappa)$ the average cavity photon number is
$\bar n_{ph} =\{ \langle a^\dag a\rangle_q\}_{cl}\simeq
\{|\alpha(t)|^2\}_{cl}$ and $\mathcal{N}(t)= |\alpha(t)|^2-
\{|\alpha(t)|^2\}_{cl}$. Under stationary conditions
$C_{\mathcal{N}}(\tau)$ can be written as
\begin{equation}\label{eq:NN}
\begin{split}
&C_{\mathcal{N}}(\tau) = \int_{0}^\infty dy_1dy_2 \,e^{-2\kappa y_1}
e^{-2\kappa y_2} \\ &
 \times \int_{-2y_1}^{2y_1} dx_1 \int_{-2y_2}^{2y_2} dx_2\, e^{-i\Delta
x_1 } \, e^{-i\Delta x_2 } \,C_4\left(\tau\!
-\!(y_1\!-\!y_2),x_1,x_2\right).
\end{split}%
\end{equation}
Here we have introduced the four point field correlation function
\begin{equation}
\begin{split}
C_4(T,t_1,t_2) =& \{
\mathcal{E}^*\left(T+t_1/2\right)\mathcal{E}\left(T-t_1/2\right)
\mathcal{E}^*\left(t_2/2\right)\mathcal{E}\left(-t_2/2\right)\}_{cl}\\
&-\{ \mathcal{E}^*\left(t_1/2\right)\mathcal{E}\left(-t_1/2\right)\}
\{
\mathcal{E}^*\left(t_2/2\right)\mathcal{E}\left(-t_2/2\right)\}_{cl}.
\end{split}
\end{equation}
For phase noise with Gaussian statistics the four point correlation
function is given by
\begin{equation}\label{eq:B3}
C_4(T,t_1,t_2)= |\mathcal{E}_0|^4 \left(e^{- \frac{1}{2}\{
\Phi^2(T,t_1,t_2)\}_{cl} }- e^{-  \frac{1}{2}\{ \phi^2(t_1)+
\phi^2(t_2)\}_{cl} }\right) \,,
\end{equation}
where $\Phi(\tau,t_1,t_2)= \int_{\tau-t_1/2}^{\tau+t_1/2} ds \,\dot
\phi(s) + \int_{-t_2/2}^{+t_2/2} ds \,\dot \phi(s)$ and
$\phi(t)=\int_{-t/2}^{+t/2} ds  \,\dot \phi(s)$.  Following the same
argumentation as in the evaluation of $S_A(\omega)$ in
App.~\ref{app:A} we can in the limit $\Gamma_L\ll \kappa$ expand the
exponentials in Eq.~\eqref{eq:B3} to first order,
\begin{equation}
C_4(T,t_1,t_2)\simeq - |\mathcal{E}_0|^4  \int_{\tau-t_1}^{\tau+t_1}
ds   \int_{-t_2}^{+t_2} ds'  \{ \dot \phi(s) \dot \phi(s')\}_{cl},
\end{equation}
or in terms of the noise spectrum  $S_{\dot \phi}(\Omega)$,
\begin{equation}
C_4(T,t_1,t_2)\simeq\! - \!4 |\mathcal{E}_0|^4   \!\int\!
\frac{d\Omega}{2\pi} S_{\dot \phi}(\Omega) \frac{\sin(\Omega
t_1/2)\sin(\Omega t_2/2)}{\Omega^2} e^{-i\Omega T} .
\end{equation}
After inserting this expression back into Eq.~\eqref{eq:NN} and
evaluating the remaining integrals we obtain
 \begin{equation}
C_{\mathcal{N}}(\tau)\simeq  |\alpha_0|^4 \int \frac{d\Omega}{2\pi}
\frac{  4 \Delta^2 S_{\dot \phi}(\Omega) e^{-i\Omega \tau}}{
(\Delta^4+2\Delta^2(\kappa^2-\Omega^2)+(\kappa^2+\Omega^2)^2)}.
\end{equation}
For the evaluation of $S_N(\omega_m)$ defined in Eq.~\eqref{eq:SN}
the integral over $\tau$ results in a term
$\sim\delta(\Omega-\omega_m)$ and we end up with
\begin{equation}
S_N(\omega_m)=g_0^2  |\alpha_0|^4 \frac{ 4\Delta^2 S_{\dot
\phi}(\omega_m)}{
(\Delta^4+2\Delta^2(\kappa^2-\omega_m^2)+(\kappa^2+\omega_m^2)^2)}.
\end{equation}
For the optimal detuning $\Delta=-\sqrt{\omega_m^2+\kappa^2}$ this
expression simplifies to $S_N(\omega_m)=g_0^2  |\alpha_0|^4 S_{\dot
\phi}(\omega_m)/\kappa^2$.

In the strong coupling regime heating rates depend on the
fluctuation spectrum evaluated at the imaginary frequency,
$S_N(\omega_\pm+i\kappa/2)$. In this case we obtain
\begin{equation}
\begin{split}
S_N(\omega_\pm+i\kappa/2)= & g_0^2  |\alpha_0|^4  \int \frac{d\Omega}{2\pi} \,  S_{\dot \phi}(\Omega)\,\frac{4 \kappa}{\kappa^2+4(\Omega-\omega_\pm)^2}  \\
&\times \frac{ 4\Delta^2 }{
(\Delta^4+2\Delta^2(\kappa^2-\Omega^2)+(\kappa^2+\Omega^2)^2)}.
\end{split}
\end{equation}
For $\Delta=-\omega_m$ and $\kappa\ll G$ this integral is dominated
by two contributions from the resonances at $\Omega=\omega_m$ and
$\Omega=\omega_\pm$. In the limit $\kappa\rightarrow 0$ we obtain
the result presented in Eq.~\eqref{eq:SNstrong}.

\section{Low frequency noise}\label{app:C}
In the absence of intensity fluctuations, $\mathcal{N}(t)=0$, the
formal solution of Eq.~\eqref{eq:ApmEvo} is given by
\begin{equation}\label{eq:C1}
\left(
\begin{matrix}
 A_{+} (t) \\
A_{-} (t)%
\end{matrix}%
\right) =e^{-i\mathbf{M}_{0}t}\left( \mathcal{T}e^{-i\int_{0}^{t}\mathbf{M}%
(s)ds}\right) \left(
\begin{matrix}
 A_{+} (0) \\
 A_{-} (0)%
\end{matrix}%
\right) ,
\end{equation}%
where $\mathcal{T}$ is the time ordering operator,
\begin{equation}
\mathbf{M}_{0}=\left(
\begin{matrix}
\omega _{+}-i\kappa /2 & 0 \\
0 & \omega _{-}-i\kappa /2%
\end{matrix}%
\right) ,
\end{equation}%
and
\begin{equation}
\mathbf{M}(t)=e^{i\mathbf{M}_{0}t}\left(
\begin{matrix}
\delta \omega (t) & -\dot{\theta}(t) \\
-\dot{\theta}(t) & -\delta \omega (t)%
\end{matrix}%
\right) e^{-i\mathbf{M}_{0}t}.
\end{equation}%
To obtain $\{c_{aa}(t)\}_{cl}$ we take the classical average of
Eq.~\eqref{eq:C1} and evaluate the average of the exponential of
$\mathbf{M}(t)$ using a second order cumulant expansion. Then
\begin{equation}\label{eq:C4}
\left(
\begin{matrix}
\{ A_{+} (t)\}_{cl} \\
\{A_{-} (t)\}_{cl}%
\end{matrix}%
\right) =e^{-i\mathbf{M}_{0}t}e^{- \bar{\bf M}(t)}\left(
\begin{matrix}
 A_{+} (0) \\
 A_{-} (0)%
\end{matrix}%
\right) ,
\end{equation}
where
\begin{equation}
\bar{\mathbf{M}}(t)=\frac{1}{2}\left(
\begin{matrix}
W(t) + R(t) + i I(t) & -X^*(t) \\
X(t) &W(t) + R(t) - i I(t)%
\end{matrix}%
\right).
\end{equation}%
Here $W(t)$ and $R(t)$  are defined in Eq.~\eqref{eq:W}, and Eq.~\eqref{eq:R}, and $I(t)$ is the imaginary part of integral in Eq.~\eqref{eq:R} which leads to a small shift of the oscillation frequency.
Finally,
\begin{equation}
\begin{split}
X(t)=2\int_0^tds \int_0^s ds'\, \Big(& e^{-i |G| s'} \{\delta\omega(s)\dot\theta(s')\}_{cl}\\
  & - e^{-i |G| s}\{\delta\omega(s')\dot\theta(s)\}_{cl}\Big),
  \end{split}
\end{equation}
 is an additional cross term which arises from correlations between $\delta \omega(t)$ and $\dot \theta(t)$. However, using the a rough approximation for $\alpha(t)\approx \alpha_0(1-i\dot \phi(t)/(\kappa+i\omega_m))$ and $\theta(t)\approx \phi(t)$ we obtain $\{\delta \omega(t) \dot \theta(t')\}_{cl}\approx |G|/\omega_m \{\dot \phi(t) \dot\phi(t')\}_{cl}$ and also this term does not contribute significantly to decoherence.
For a simplified discussion, both the cross term and the frequency shift are omitted in Eq.~\eqref{eq:FidelityLowFreq}.

\end{appendix}


\begin{thebibliography}{99}
\bibitem{Thermalnoise} Our analysis can be extended to thermal fields in a
straight-forward way by using the thermal expectation values
$\langle a^\dag a\rangle_q=N_{th}$ and $\langle a a^\dag
\rangle_q=N_{th}+1$ in the perturbative derivation of
Eq.~\eqref{eq:cooling}. However, the main effect of thermal noise is
simply an additional limitation $n_0 \geq N_{th}$ which is
independent of the laser noise and intensity and does not change the
main conclusions of this paper.
\bibitem{Average} By taking the classical average we obtain Eq.~%
\eqref{eq:cooling} plus additional corrections of order $2g_0/
\kappa \{ \mathcal{N}(t)\langle b^\dag b\rangle_q(t)\}_{cl}$. This
correlator is not
necessarily zero but in the limit of small fluctuations, $|\mathcal{N}%
(t)|\ll n_{ph} $, it does not considerably modify the dynamics.
\end{thebibliography}
\end{document}